\documentclass[twocolumn]{autart}    

\usepackage{graphicx}          

\usepackage{amsmath}
\usepackage{amsfonts}
\usepackage[nolist]{acronym}    
\usepackage{amssymb} 

\usepackage{natbib} 

\usepackage{empheq} 

\usepackage{tikz,ifthen}
\usetikzlibrary{calc,decorations.pathreplacing,calligraphy}

\newcommand{\eqnref}[1]{(\ref{#1})}

\newcommand{\secref}[1]{Section~\ref{#1}}
\newcommand{\inv}{^{\raisebox{.2ex}{$\scriptscriptstyle-1$}}}

\newcommand{\tHf}{\widetilde{\mathcal{H}}_f}

\newcommand{\tKp}[1]{\widetilde{\mathcal{K}}_{p}^{\mathrm{{#1}}}}

\newcommand{\datavec}[2]{\underline{#1}_{#2}}

\begin{document}

\begin{frontmatter}

\title{Closed-loop Data-Enabled Predictive Control and its equivalence with Closed-loop Subspace Predictive Control\thanksref{footnoteinfo}} 

\thanks[footnoteinfo]{This paper was not presented at any IFAC meeting.}

\author[TUD]{R. Dinkla\corauthref{cor}}\ead{r.t.o.dinkla@tudelft.nl},    
\corauth[cor]{Corresponding author.%
}
\author[TUD]{S. P. Mulders}\ead{s.p.mulders@tudelft.nl},               
\author[TUD,TUE]{T. Oomen}\ead{t.a.e.oomen@tudelft.nl},  
\author[TUD]{J.W. van Wingerden}\ead{j.w.vanwingerden@tudelft.nl}
\address[TUD]{Delft Center for Systems and Control, Delft University of Technology, Mekelweg 2, 2628CD Delft, The Netherlands}  
\address[TUE]{Control Systems Technology Group, Eindhoven University of Technology, 5600
MB Eindhoven, The Netherlands}        


\begin{abstract}
Factors like improved data availability and increasing system complexity have sparked interest in data-driven predictive control (DDPC) methods like Data-enabled Predictive Control (DeePC). However, closed-loop identification bias arises in the presence of noise, which reduces the effectiveness of obtained control policies. In this paper we propose Closed-loop Data-enabled Predictive Control (CL-DeePC), a framework that unifies different approaches to address this challenge. To this end, CL-DeePC incorporates instrumental variables (IVs) to synthesize and sequentially apply consistent single or multi-step-ahead predictors. Furthermore, a computationally efficient CL-DeePC implementation is developed that reveals an equivalence with Closed-loop Subspace Predictive Control (CL-SPC). Compared to DeePC, CL-DeePC simulations demonstrate superior reference tracking, with a sensitivity study finding a 48\% lower susceptibility to noise-induced reference tracking performance degradation.
\end{abstract}

\end{frontmatter}
\begin{acronym}%
    \acro{SPC}{Subspace Predictive Control}
    \acro{CL-SPC}{Closed-loop Subspace Predictive Control}
    \acro{DeePC}{Data-enabled Predictive Control}
    \acro{LTI}{linear time-invariant}
    \acro{IV}{instrumental variable}
    \acro{IVs}{instrumental variables}
    \acro{N4SID}{Numerical algorithm for Subspace State Space System Identification}
    \acro{MPC}{Model Predictive Control}
    \acro{CL-DeePC}{Closed-loop Data-enabled Predictive Control}
    \acro{DDPC}{Data-driven Predictive Control}
\end{acronym}%
\section{Introduction}\label{sec:introduction}
\noindent Trends of increasing data availability and system complexity provide opportunities for data driven-driven control~\citep{Hou2013}. In sharp contrast to the use of data in indirect data-driven approaches to synthesise a model by means of system identification, direct data-driven control approaches are promising because of their ability to derive a control law directly from data without having to realize an explicit system model as an often expensive intermediate step~\citep{Hjalmarsson2005}.

A direct \ac{DDPC} method called \acf{DeePC} is developed in~\cite{Coulson2019} and has recently seen considerable development and successful implementation (see, e.g., \citet{Markovsky2023} and~\citet{Breschi2023a}). \ac{DeePC} relies on Willems' Fundamental Lemma in a receding horizon optimal control framework. This lemma reveals that for a deterministic \ac{LTI} system, any sufficiently persistently exciting input-output trajectory parameterizes all possible future input-output trajectories~\citep{Willems2005}.

For non-deterministic systems, care has to be taken with such a parameterization in terms of only past input-output trajectories because this does not consider the effects of noise. If the block-Hankel data matrix in which the noisy input-output trajectories are stored is full rank then unattainable future trajectories may be predicted~\citep{Markovsky2023}. On the other hand, if the data matrix is rank-deficient, then the \ac{DeePC} problem may be infeasible. To deal with noise, slack variables and regularization initially served as heuristic measures to introduce robustness~\citep{Coulson2019}, and subsequently have been motivated formally to, e.g., provide robust closed-loop stability guarantees~\citep{Berberich2021}, distributional robustness~\citep{Coulson2019a}, and robustness to structured uncertainty~\citep{Huang2023}. Other approaches to handle noise include averaging techniques~\citep{Sassella2022a}, singular value based thresholding~\citep{Sassella2022}, 
and the use of maximum likelihood estimation~\citep{Yin2023}. See also~\cite{Sassella2023} for a discussion of such methods, or \cite{Verheijen2023} for a practical review of several \ac{DDPC} methods.

To fundamentally address the consequences of noise in a data-driven setting, an \ac{IV} approach is presented in~\cite{vanWingerden2022}. Using \acs{IVs} to mitigate noise, this method demonstrates the equivalence between \ac{DeePC} and a subspace identification-inspired direct \ac{DDPC} method called \ac{SPC} from~\cite{Favoreel1999}\footnote{This equivalence is also shown by~\cite{Fiedler2021} using regularizations and in a noiseless setting.}. The established equivalence provides opportunities for in-depth analysis and further development of the direct \ac{DDPC} techniques by the strong fundamental basis of subspace identification methods.

The correlation between inputs and noise that arises from feedback results in closed-loop identification bias~\citep{Ljung1996}. \cite{Dinkla2023} demonstrate that this problem can arise with batch-wise adaptive\footnote{In batch-wise adaptive operation, subsequent controllers employ closed-loop data from only a single controller. This paper considers (fully) adaptive implementations whereby the controller is updated at each time step.} \ac{SPC} and (given the aforementioned equivalence also) \ac{DeePC} applications, potentially degrading controller performance. To tackle the closed-loop identification issue~\cite{Dinkla2023} suggest using \ac{IVs} and sequential step-ahead predictions, the use of which is further confirmed by work of~\cite{Wang2023} and~\cite{Shi2023}. Drawing on subspace identification methods, the idea of sequential step-ahead predictions is also used by \acf{CL-SPC}~\citep{Dong2008}.

Although data-driven control algorithms have seen considerable development, to date, optimal noise mitigation under feedback is not completely addressed.
The aim of this paper is to rigorously establish \ac{CL-DeePC}, thereby providing a unifying framework of different solution strategies that addresses closed-loop identification bias.

The main contributions of this paper are: 
\begin{enumerate}%
\item[1.] to formally establish \acf{CL-DeePC} as a new \ac{DDPC} framework, \label{contribution:develop_CL_DeePC}
\item[2.] to incorporate \ac{IVs} in \ac{CL-DeePC} as a systematic noise-mitigation technique to provide consistent and causal single-step-ahead predictors or consistent multiple-step-ahead predictors, \label{contribution:incorporate_IVs}
\item[3.] to present a unified \ac{CL-DeePC} framework that solves the closed-loop identification problem that arises in the presence of feedback and noise by sequential application of such consistent predictors,\label{contribution:solves_CL_issue}
\item[4.] to present a computationally efficient \ac{CL-DeePC} technique that reveals an equivalence between \ac{CL-DeePC} and \ac{CL-SPC},
\item[5.] to show the superior performance of \ac{CL-DeePC} compared to \ac{DeePC} in simulation.
\end{enumerate}

This paper is structured as follows. \secref{sec:prelim} introduces the used system model and notation. In \secref{sec:CL-DeePC} \ac{CL-DeePC} is developed using \ac{IVs} to obtain a unified \ac{DDPC} framework that encompasses both consistent sequential single and multi-step-ahead predictions. \secref{sec:Sequential} develops a computationally efficient \ac{CL-DeePC} implementation for which \secref{sec:equivalence2CLSPC} subsequently reveals an equivalence with \ac{CL-SPC}. \secref{sec:results} presents simulation results that facilitate a comparison between the performance of \ac{DeePC} and \ac{CL-DeePC}. Finally, conclusions are presented in \secref{sec:conclusion}.
\section{Preliminaries}\label{sec:prelim}
\noindent This section presents the employed system model, notation, and the considered control problem.

\subsection{System model}\label{sec:sys_model}
\noindent Consider a non-deterministic discrete \ac{LTI} system $\mathcal{S}$ whose dynamics is described in the so-called innovation form by
\begin{subequations}\label{eqn:SS_innovation}
\begin{empheq}[left=\mathcal{S}_\mathcal{I}\empheqlbrace]{align}
    x_{k+1} &= Ax_k + Bu_k + Ke_k,\label{eqn:SSi_x}\\
	y_k &= Cx_k + Du_k + e_k \label{eqn:SSi_y},
  \end{empheq}
\end{subequations}
in which the subscript $k\in\mathbb{Z}$ denotes the discrete time index, ${x_k\in\mathbb{R}^n}$, ${u_k\in\mathbb{R}^r}$, ${y_k\in\mathbb{R}^l}$, ${e_k\in\mathbb{R}^l}$ respectively represent states, inputs, outputs, and zero-mean white innovation noise with variance $\Sigma(e_k) \succ 0$, and $\{A,B,C,D,K\}$ are system matrices of compatible dimensions. It is assumed that the input and output sequences of $\mathcal{S}$ are (jointly) quasi-stationary second-order ergodic stochastic processes. This ensures that limits of time averages involving these sequences exist and that sample correlations approach a finite true correlation with probability one as the number of samples goes to infinity~\citep{Ljung1999}. Without loss of generality it is assumed that the data is generated by a minimal system realization. Moreover, $K$ represents a Kalman filter gain matrix that renders ${\tilde{A}=A-KC}$ asymptotically stable such that $\rho=\max|\text{eig}(\tilde{A})|<1$ (see, e.g., \citet[Sec.~5.7]{Verhaegen2007a}). Rearranging and substituting \eqnref{eqn:SSi_y} into \eqnref{eqn:SSi_x} obtains the equivalent predictor form
\begin{subequations}\label{eqn:SS_predictor}
\begin{empheq}[left=\mathcal{S}_\mathcal{P}\empheqlbrace]{align}
	x_{k+1} &= \tilde{A}x_k + \tilde{B}u_k + Ky_k,\label{eqn:SSp_x}\\
	y_k &= Cx_k + Du_k + e_k \label{eqn:SSp_y},
  \end{empheq}
\end{subequations}
in which $\tilde{B}=B-KD$.
\subsection{Notation and definitions}\label{sec:notation}
\noindent This section introduces useful notation and definitions.
Several block-Toeplitz matrices are defined by
\begin{align}\label{eqn:blockToeplitz} 
\mathcal{T}_s(\mathcal{A},\mathcal{B},\mathcal{C},\mathcal{D}) =\scriptsize{
	\begingroup
    \renewcommand*{\arraystretch}{0.8}
    \begin{bmatrix}
		\mathcal{D}         & 0         & 0      & \cdots  & 0\\
		\mathcal{C}\mathcal{B}        & \mathcal{D}         & 0      & \cdots  & 0\\
		\mathcal{C}\mathcal{A}\mathcal{B}       & \mathcal{C}\mathcal{B}        & \mathcal{D}      & \cdots & 0\\
		\vdots    &  \vdots & \ddots & \ddots & \vdots\\
		\mathcal{C}\mathcal{A}^{s-2}\mathcal{B} & \mathcal{C}\mathcal{A}^{s-3}\mathcal{B} & \cdots  & \mathcal{C}\mathcal{B}     & \mathcal{D}
	\end{bmatrix}
    \endgroup},
\end{align}
in which the matrices $\{\mathcal{A},\mathcal{B},\mathcal{C},\mathcal{D}\}$ are of compatible dimensions. Let $s\in\mathbb{Z}_{>0}$ denote a generic strictly-positive integer. As a subscript, $s$ here indicates the number of block-rows. Let ${I_s\in\mathbb{R}^{s\times s}}$ represent an identity matrix. Equation~\eqnref{eqn:blockToeplitz} then defines the block-Toeplitz matrices
\begin{alignat*}{2}
\mathcal{T}_s^\mathrm{u}&=\mathcal{T}_s(A,B,C,D),\quad  &\mathcal{H}_s&=\mathcal{T}_s(A,K,C,I_l),\\
\widetilde{\mathcal{T}}_s^\mathrm{u}&=\mathcal{T}_s(\tilde{A},\tilde{B},C,D),\quad  &\widetilde{\mathcal{H}}_s&=\mathcal{T}_s(\tilde{A},K,-C,I_l).
\end{alignat*}

In addition, for a generic $s$ and specific past data window length $p\in\mathbb{Z}_{>0}$, the extended observability matrices $\Gamma_s$ and $\widetilde{\Gamma}_s$ as well as extended reversed controllability matrices $\tKp{u}$, $\tKp{y}$ are defined by
$$\begin{array}{rll}
\Gamma_s &= \begin{bmatrix}C^\top & (CA)^\top & \cdots & (CA^{s-1})^\top\end{bmatrix}^\top&\in\mathbb{R}^{sl\times n},\\
\widetilde{\Gamma}_s &= \begin{bmatrix}C^\top & (C\tilde{A})^\top & \cdots & (C\tilde{A}^{s-1})^\top\end{bmatrix}^\top&\in\mathbb{R}^{sl\times n},\\
\tKp{u} &= \begin{bmatrix} \tilde{A}^{p-1}\tilde{B}\, & \tilde{A}^{p-2}\tilde{B} & \cdots & \tilde{A}\tilde{B} & \tilde{B}\,\,\end{bmatrix}&\in\mathbb{R}^{n\times pr},\vphantom{\bigg|}\\
\tKp{y} &= \begin{bmatrix} \tilde{A}^{p-1}K & \tilde{A}^{p-2}K & \cdots & \tilde{A}K & K \end{bmatrix}&\in\mathbb{R}^{n\times pl}.
\end{array}$$
\noindent Data vectors are denoted as
\begin{align*}
    \datavec{u}{k,s} = \begin{bmatrix} u_k^\top & u_{k+1}^\top & \cdots & u_{k+s-1}^\top\end{bmatrix}^\top\in\mathbb{R}^{sr},
\end{align*}
which represents a vector of ordered input data starting at time index $k$, and containing a number of samples $s$.

Using such data vectors, it is possible to concisely define block-Hankel data matrices. Such a block-Hankel data matrix is given by
\begin{align*}
    U_{k,s,q} = \frac{1}{\sqrt{q}}\begin{bmatrix}
        \datavec{u}{k,s} & \datavec{u}{k+1,s} & \cdots & \datavec{u}{k+q-1,s}
    \end{bmatrix}\in\mathbb{R}^{sr\times q},
\end{align*}
in which $q\in\mathbb{Z}_{>0}$ is another generic positive-definite integer that here represents the number of successive input data vectors with $s$ data samples each, starting from time index $k$. Note the block-anti-diagonal structure of block-Hankel matrices. We indicate vectors and matrices that are composed entirely or partly of predictions by respectively $\hat{(\cdot)}$ or $\overline{(\cdot)}$. 

Block-Hankel data matrices are employed to define the notion of persistency of excition below, with ${N\in\mathbb{Z}_{>0}}$ as a number of columns.
\begin{defn}\label{def:PE}\citep[Def.~10.1]{Verhaegen2007a} A signal consisting of samples ${w_j\!\in\!\mathbb{R}^q},$ $j\!\in\![k,k+s+N-2]$ is persistently exciting of order $s$ if the associated block-Hankel matrix ${ W_{k,s,N}\in\mathbb{R}^{sq \times N}}$ is full row rank.
\end{defn}

For notational convenience, $\Psi$ denotes a concatenation of input and output block-Hankel matrices:
\begin{align}\label{eq:Psi_def}
    \Psi_{k,s,q} = \begin{bmatrix}
        U_{k,p,q}^\top & U_{k_p,s,q}^\top & Y_{k,p,q}^\top
    \end{bmatrix}^\top\in\mathbb{R}^{((p+s)r+pl)\times q},
\end{align}
in which $k$ indicates the starting index, and $s$ and $q$ parameterize the dimensions of the concatenated matrix together with $p$.

Furthermore, the expectation of a variable with a stochastic component is denoted by ${\mathbb{E}[\cdot]}$. For sequences of scalar random variables $a_N$ and $b_N$, ${a_N = o(b_N)}$ indicates that ${\lim_{N\rightarrow\infty}a_N/b_N=0}$. Likewise ${a_N=o_P(b_N)}$ stipulates that $\forall \delta > 0$, ${\lim_{N\rightarrow\infty}P[|a_N/b_N|>\delta]=0}$.

In addition, the well-posedness of a closed-loop system is defined as follows.
\begin{defn}\label{def:well-posedness}\citep{VanOverschee1997}
    A closed-loop system composed of an \ac{LTI} plant and controller with respectively direct feedthrough matrices $D$ and $D_\mathrm{c}$ is well-posed if $I_l+DD_\mathrm{c}$ is invertible.
\end{defn}
The outputs of a well-posed closed-loop system are uniquely defined by the reference and the states of the plant and controller. Note that a practical sufficient condition for well-posedness is that either $D=0$ or $D_\mathrm{c}=0$.
%
\subsection{The data equations}\label{sec:DerivingDataEquations}
\noindent This section derives several fundamental relations called data-equations, which reformulate \eqref{eqn:SS_innovation} and \eqref{eqn:SS_predictor} in terms of block-Hankel matrices.

To this end, iterative application of respectively \eqref{eqn:SS_innovation} and \eqref{eqn:SS_predictor} leads to%
\begin{align}
    Y_{k_p,s,q} &= \Gamma_s X_{k_p,1,q} + \mathcal{T}_s^\mathrm{u} U_{k_p,s,q} + \mathcal{H}_s E_{k_p,s,q}\label{eq:Yf1},\\
    \begin{split}%
    Y_{k_p,s,q} &= \widetilde{\Gamma}_s X_{k_p,1,q} + \widetilde{\mathcal{T}}_s^\mathrm{u} U_{k_p,s,q} + E_{k_p,s,q}\\
    &\phantom{=}+(I_{sl}-\widetilde{\mathcal{H}}_s)Y_{k_p,s,q},
    \end{split}\label{eq:Yf2}
\end{align}
in which we employ a recurring shorthand for time indices exemplified by $k_p=k+p$. Furthermore, the initial states can be rewritten in terms of preceding states and input-output data using \eqref{eqn:SS_predictor} as%
\begin{align}\label{eq:Xip}
    X_{k_p,1,q} = \tilde{A}^p X_{k,1,q} + \tKp{u} U_{k,p,q} + \tKp{y} Y_{k,p,q}.
\end{align}
Substitute \eqref{eq:Xip} into \eqref{eq:Yf1} and \eqref{eq:Yf2} 
to find the \textit{data equations}
\begin{align}
    Y_{k_p,s,q} &= L_s \Psi_{k,s,q} + \mathcal{H}_s E_{k_p,s,q} + \Gamma_s \tilde{A}^p X_{k,1,q},\label{eq:DataEq1}\\
    \begin{split}
    Y_{k_p,s,q} &= \widetilde{L}_s \Psi_{k,s,q} + E_{k_p,s,q} + (I_{sl}-\widetilde{\mathcal{H}}_s) Y_{k_p,s,q} \\
    &\phantom{=}+ \widetilde{\Gamma}_s \tilde{A}^p X_{k,1,q},
    \end{split}\label{eq:DataEq2}
\end{align}
in which $L_s,\widetilde{L}_s\in\mathbb{R}^{sl\times ((p+s)r+pl)}$ are defined as
\begin{align*}
    L_s &= \begin{bmatrix} \Gamma_s\tKp{u} & \mathcal{T}_s^\mathrm{u} & \Gamma_s\tKp{y} \end{bmatrix},
    \quad \widetilde{L}_s = \begin{bmatrix} \widetilde{\Gamma}_s\tKp{u} & \widetilde{\mathcal{T}}_s^\mathrm{u} & \widetilde{\Gamma}_s\tKp{y} \end{bmatrix}
    .
\end{align*}

\subsection{Receding horizon control problem formulation}
\noindent This section clarifies the scope of this work by considering a receding horizon control problem, exemplified by
\begin{subequations}
\begin{alignat}{2}
    \span\span\min_{\datavec{u}{\hat{i}_p,f}} ||\datavec{\hat{y}}{\hat{i}_p,f}-\datavec{r}{\hat{i}_p,f}||_Q^2 + ||\datavec{u}{\hat{i}_p,f}||_R^2 + ||\Delta\datavec{u}{\hat{i}_p,f}||_{R^\Delta}^2 \span\label{eq:cost_fun}\\
    \text{s.t.}\quad& &\datavec{\hat{y}}{\hat{i}_p,f}&=\Gamma_f x_{\hat{i}_p}+\mathcal{T}_f^\mathrm{u}\datavec{u}{\hat{i}_p,f},\label{eq:SS_iter}\\
   && x_{\hat{i}_p}&=x_\mathrm{ini},\label{eq:x_ini}\\
   && u_k&\in\mathcal{U},\quad \hat{y}_k\in\mathcal{Y},\quad \forall k\in\big[\hat{i}_p,\,\hat{i}_p+f\big),
\end{alignat}
\end{subequations}
in which $\datavec{\hat{y}}{\hat{i}_p,f}$ and $\datavec{r}{\hat{i}_p,f}$ are vectors of respectively output predictions and a future reference trajectory, $x_\mathrm{ini}$ specifies an initial state by~\eqref{eq:x_ini} for the predictive model~\eqref{eq:SS_iter}, and $||\cdot||_{(\cdot)}$ denotes a weighted Euclidian norm. Moreover, ${\Delta\datavec{u}{\hat{i}_p,f}=\datavec{u}{\hat{i}_p,f}-\datavec{u}{\hat{i}_p-1,f}}$, $Q\in\mathbb{R}^{fl\times fl}$ is a positive semi-definite weighting matrix, and $R,R^\Delta\in\mathbb{R}^{fr\times fr}$ are positive definite positive definite weighting matrices. Furthermore, $\mathcal{U}$ and $\mathcal{Y}$ respectively represent sets of allowable inputs and outputs.

Without knowledge of the system matrices $\{A,B,C,D,K\}$ and the initial state $x_\mathrm{ini}$, but given sufficiently informative past input-output data from  intervals $k\in[i,\,i+\bar{N})$ and $k\in[\hat{i},\,\hat{i}_p)$ that may overlap\footnote{Depending on the difference $\hat{i}-i>0$ and the number of samples $\bar{N}$. For fully adaptive implementations $i+\bar{N}=\hat{i}_p$.} and collected in closed-loop, this paper presents a unified framework to synthesize a consistent data-driven output predictor with which to replace the unknown model-based relations \eqref{eq:SS_iter} and \eqref{eq:x_ini}.
\section{Closed-loop Data-enabled Predictive Control}\label{sec:CL-DeePC}
\noindent In this section, \ac{CL-DeePC} is established, thereby providing Contribution~1. The fundamental idea of sequential single-step-ahead predictors is explained first before a rigorous exposition introduces \ac{IVs} to assure the consistency of these predictors. Results are subsequently generalized to the sequential application of multi-step-ahead predictors, thereby providing Contributions~2 and~3.

\subsection{Fundamental idea: sequential \ac{DeePC} with $f=1$}
\begin{figure}[b!]
\centering
\input{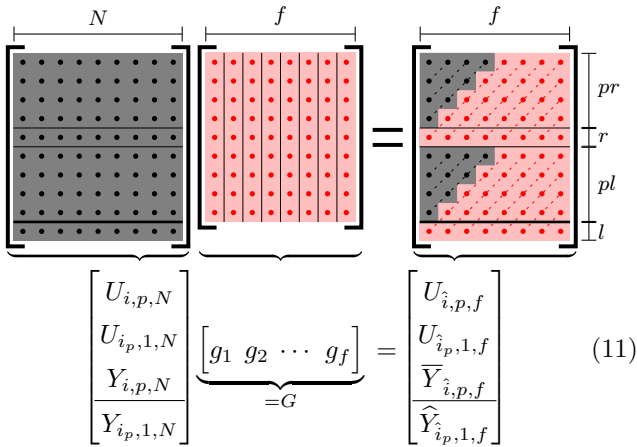}
\caption{Visualization of known (black) and unknown (red) variables in \ac{CL-DeePC} without \ac{IVs}. The composition of block-Hankel matrices on the right hand side (note the dashed anti-diagonals) results from sequential application of \ac{DeePC} with $f=1$.}
\label{fig:CL-DeePC}
\end{figure}
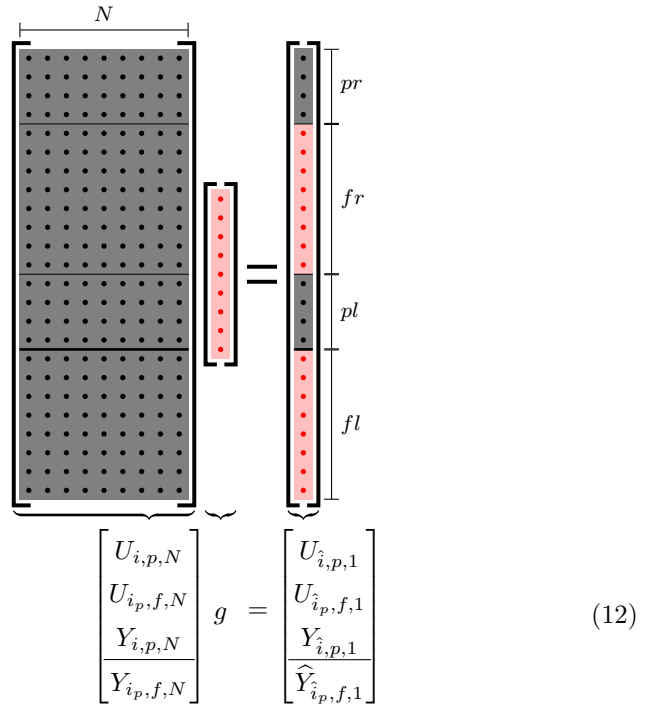
\begin{figure}[b!]
\centering
\begin{tikzpicture}
    \def\stepSize{0.25}
    \def\Nnum{9}
    \def\fnum{8}
    \def\pnum{4}
    
    \setlength{\onelen}{\stepSize cm}
    \setlength{\BrCl}{0.075cm}
    \setlength{\BrIn}{0.15cm}
    \setlength{\plen}{1cm}
    \setlength{\flen}{2cm}
    \setlength{\Nlen}{2.25cm}
    \setlength{\MatClearance}{0.3cm}

    \path (0,2\plen+\onelen) coordinate (M1D);
    \path ([yshift=-2\plen-2\flen]M1D) coordinate (M1A);
    \path ([xshift=\Nlen]M1A) coordinate (M1B);
    \path ([yshift=2*\plen+2*\flen]M1B) coordinate (M1C);
    \draw[line width=1.5pt] ([xshift=\BrIn,yshift=-\BrCl]M1A) -- ([xshift=-\BrCl,yshift=-\BrCl]M1A) -- ([xshift=-\BrCl,yshift=\BrCl]M1D) -- ([xshift=\BrIn,yshift=\BrCl]M1D); 
    \draw[line width=1.5pt] ([xshift=-\BrIn,yshift=-\BrCl]M1B) -- ([xshift=\BrCl,yshift=-\BrCl]M1B) -- ([xshift=\BrCl,yshift=\BrCl]M1C) -- ([xshift=-\BrIn,yshift=\BrCl]M1C); 
    \draw[line width=1pt] ([yshift=\flen]M1A) -- ([yshift=\flen]M1B); 
    \fill[black, opacity=0.5] (M1A) rectangle (M1C);
    \foreach \x in {0,...,8} { 
    \foreach \y in {-15,...,8} {
      \fill ( {(\x+0.5)*\onelen}, {(\y+0.5)*\onelen} ) circle (1pt);
    }}
    \draw[line width=0.1pt] ([yshift=-\plen-\flen]M1D) -- ([yshift=-\plen-\flen]M1C);
    \draw[line width=0.1pt] ([yshift=-\plen]M1D) -- ([yshift=-\plen]M1C);

    \path ([yshift=-\plen]M1D) coordinate (M1stair1A);
    \path ([xshift=\Nlen]M1D) coordinate (M1stair1I);
    
    \path ([xshift=-0.5\BrIn,yshift=-2\BrCl]M1A) coordinate (Brace1L);
    \path ([xshift=0.5\BrIn,yshift=-2\BrCl]M1B) coordinate (Brace1R);
    \draw[decorate, decoration={calligraphic brace, amplitude=3pt, mirror, aspect=0.75},line width=1pt] (Brace1L) -- (Brace1R);
    \node (mat1) at ($(Brace1L)!0.75!(Brace1R) + (0,-3pt)$) {};
    \node[below] at (mat1.center) {$\begin{bmatrix}
        U_{i,p,N}\\U_{i_p,f,N}\\Y_{i,p,N}\\ \hline Y_{i_p,f,N}
    \end{bmatrix}$};
  
    \path ([xshift=\MatClearance,yshift=-0.5\Nlen-\plen-\flen]M1C) coordinate (M2A);
    \path ([xshift=\onelen]M2A) coordinate (M2B);
    \path ([yshift=\Nlen]M2B) coordinate (M2C);
    \path ([xshift=-\onelen]M2C) coordinate (M2D);

    \draw[line width=1.5pt] ([xshift=0.5\BrIn,yshift=-\BrCl]M2A) -- ([xshift=-\BrCl,yshift=-\BrCl]M2A) -- ([xshift=-\BrCl,yshift=\BrCl]M2D) -- ([xshift=0.5\BrIn,yshift=\BrCl]M2D); 
    \draw[line width=1.5pt] ([xshift=-0.5\BrIn,yshift=-\BrCl]M2B) -- ([xshift=\BrCl,yshift=-\BrCl]M2B) -- ([xshift=\BrCl,yshift=\BrCl]M2C) -- ([xshift=-0.5\BrIn,yshift=\BrCl]M2C); 
    
    \fill[red!50,opacity=0.5] (M2A) rectangle (M2C);

    \foreach \x in {0} {
    \foreach \y in {0,...,8} {
      \fill[red] ([xshift=(\x+0.5)*\onelen,yshift=(\y+0.5)*\onelen]M2A) circle (1pt);
    }}

    \coordinate (Brace2L) at ([xshift=-0.5\BrIn]M2A |- Brace1L);
    \coordinate (Brace2R) at ([xshift=0.5\BrIn]M2B |- Brace1L);
    \draw[decorate, decoration={calligraphic brace, amplitude=3pt, mirror, aspect=0.5},line width=1pt] (Brace2L) -- (Brace2R);
    \node (mat1) at ($(Brace2L)!0.5!(Brace2R) + (0,-3pt)$) {};
    \node[below] at ([yshift=-1.05cm]mat1.center) {$g$};
    
    \path ([xshift=\MatClearance*3/4,yshift=\Nlen/2-0.1cm]M2B) coordinate (EqA);
    \path ([xshift=0.4cm]EqA) coordinate (EqB);
    \path ([yshift=0.2cm]EqB) coordinate (EqC);
    \path ([xshift=-0.4cm]EqC) coordinate (EqD);
    \draw[line width = 1.5 pt] (EqA) -- (EqB);
    \draw[line width = 1.5 pt] (EqD) -- (EqC);

    \node[below] at ([xshift=2\onelen,yshift=-1.05cm]mat1.center) {$=$};

    \path ([xshift=0.75\MatClearance]EqB |- M1A) coordinate (M3A);
    \path ([xshift=\onelen]M3A) coordinate (M3B);
    \path ([yshift=2\plen+2\flen]M3B) coordinate (M3C);
    \path ([xshift=-\onelen]M3C) coordinate (M3D);
    
    \fill[black,opacity=0.5]  (M3D) -- ([yshift=-\plen]M3D) -- ([yshift=-\plen]M3C) -- (M3C) -- cycle;
    \fill[red!50,opacity=0.5] ([yshift=-\plen]M3D) -- ([yshift=-\plen-\flen]M3D) -- ([yshift=-\plen-\flen]M3C) -- ([yshift=-\plen]M3C) -- cycle;
    \fill[black,opacity=0.5] ([yshift=\flen]M3A) -- ([yshift=\flen]M3B) -- ([yshift=\flen+\plen]M3B) -- ([yshift=\flen+\plen]M3A) -- cycle;
    \fill[red!50,opacity=0.5]  (M3A) -- (M3B) -- ([yshift=\flen]M3B) -- ([yshift=\flen]M3A) -- cycle;
    
    \draw[line width=1.5pt] ([xshift=0.5\BrIn,yshift=-\BrCl]M3A) -- ([xshift=-\BrCl,yshift=-\BrCl]M3A) -- ([xshift=-\BrCl,yshift=\BrCl]M3D) -- ([xshift=0.5\BrIn,yshift=\BrCl]M3D); 
    \draw[line width=1.5pt] ([xshift=-0.5\BrIn,yshift=-\BrCl]M3B) -- ([xshift=\BrCl,yshift=-\BrCl]M3B) -- ([xshift=\BrCl,yshift=\BrCl]M3C) -- ([xshift=-0.5\BrIn,yshift=\BrCl]M3C); 
    
    \foreach \dy in {0,...,-3}{
        \fill[black] ([xshift=0.5\onelen,yshift=(\dy-0.5)*\onelen]M3D) circle (1pt);
    }

    \foreach \dy in {-4,...,-11}{
        \fill[red] ([xshift=0.5\onelen,yshift=(\dy-0.5)*\onelen]M3D) circle (1pt);
    }
    
    \foreach \dy in {0,...,-3}{
        \fill[black] ([xshift=0.5\onelen,yshift=(\dy-0.5)*\onelen-\plen-\flen]M3D) circle (1pt);
    }

    \foreach \dy in {-4,...,-11}{
        \fill[red] ([xshift=0.5\onelen,yshift=(\dy-0.5)*\onelen-\plen-\flen]M3D) circle (1pt);
    }

    \draw[line width=0.1pt] ([yshift=-\plen]M3D) -- ([yshift=-\plen]M3C);
    \draw[line width=0.1pt] ([yshift=-\plen-\flen]M3D) -- ([yshift=-\plen-\flen]M3C);
    \draw[line width=1pt] ([yshift=\flen]M3A) -- ([yshift=\flen]M3B);

    \path ([xshift=-0.5\BrIn,yshift=-2\BrCl]M3A) coordinate (Brace3L);
    \path ([xshift=0.5\BrIn,yshift=-2\BrCl]M3B) coordinate (Brace3R);
    \draw[decorate, decoration={calligraphic brace, amplitude=3pt, mirror, aspect=0.5},line width=1pt] (Brace3L) -- (Brace3R);
    \node (mat3) at ($(Brace3L)!0.5!(Brace3R) + (0,-3pt)$) {};
    \node[below] at ([xshift=0.35cm]mat3.center) {$\begin{bmatrix}
        U_{\hat{i},p,1}\\U_{\hat{i}_p,f,1}\\Y_{\hat{i},p,1}\\ \hline \widehat{Y}_{\hat{i}_p,f,1}
    \end{bmatrix}$};
    \node[below left] at (8.35cm,-5cm) {$\refstepcounter{equation}(\theequation)\label{eq:regular_DeePC_no_IVs}$};
    
    \draw[|-|] ([yshift=\onelen]M1D) -- node[above] {\scriptsize$N$} ([yshift=\onelen]M1C);
    \draw[|-|] ([xshift=\onelen]M3C) -- node[right] {\scriptsize$pr$} ([xshift=\onelen,yshift=-\plen]M3C);
    \draw[|-|] ([xshift=\onelen,yshift=-\plen]M3C) -- node[right] {\scriptsize$fr$} ([xshift=\onelen,yshift=-\plen-\flen]M3C);
    \draw[|-|] ([xshift=\onelen,yshift=-\plen-\flen]M3C) -- node[right] {\scriptsize$pl$} ([xshift=\onelen,yshift=\flen]M3B);
    \draw[|-|] ([xshift=\onelen,yshift=\flen]M3B) -- node[right] {\scriptsize$fl$} ([xshift=\onelen]M3B);
\end{tikzpicture}
\caption{Visualization of known (black) and unknown (red) variables in \ac{DeePC} without \ac{IVs}. Dots represent an input $u_k\in\mathbb{R}^r$, output $y_k\in\mathbb{R}^l$, or element of the vector $g$. An $f$-step-ahead predictor is formed directly by taking a linear combination of past input and output data.}
\label{fig:regular-DeePC}
\end{figure}
\noindent A solution to the identification bias that arises in closed-loop due to correlation between inputs and noise is the synthesis of a one-step-ahead predictor\footnote{This works for practical closed-loop systems with at least one sample delay~\citep{Ljung1996}, which are consequently also well-posed by Definition~\ref{def:well-posedness}.}. For receding horizon optimal control settings with a typically more useful, larger prediction horizon ($f>1$), the described one-step-ahead predictor can be applied sequentially. \ac{CL-DeePC}, as visualized by Fig.~\ref{fig:CL-DeePC}, sequentially applies a single-step-ahead predictor that is based on \ac{DeePC}, which is shown by Fig.~\ref{fig:regular-DeePC}.

To obtain a single-step-ahead predictor for sequential use, consider \ac{DeePC} as shown by Fig.~\ref{fig:regular-DeePC} and~\eqref{eq:regular_DeePC_no_IVs}. Black dots represent known past inputs and outputs whilst red dots resemble unknown optimization variables, which are comprised of future inputs and output predictions and elements of the vector $g$. Known past data on the right hand side encodes information on an initial state, restricting the allowable set of vectors $g$ in~\eqref{eq:regular_DeePC_no_IVs}. Along the lines of Willems' Fundamental Lemma, future inputs and output predictions $\widehat{Y}_{\hat{i}_p,f,1}$ are parameterized by a linear combination $g$ of past input-output trajectories. As in~\eqref{eq:regular_DeePC_no_IVs} \ac{DeePC} thus provides a one-step-ahead predictor with $f=1$.

\ac{CL-DeePC} and its sequential use of such one-step-ahead predictors is shown by Fig.~\ref{fig:CL-DeePC} and~\eqref{eq:CL_DeePC_no_IVs}. Successive vectors $g_k$ in $G$ take linear combinations of the same past data matrix on the left hand side, but parameterize input-output trajectories on the right hand side that are successively shifted by a single time step further into the future. The sequential application of a single-step-ahead predictor is apparent from the introduction by subsequent columns on the right hand side of a single unknown new input and output prediction. This results in the composition of block-Hankel matrices on the right hand side of~\eqref{eq:CL_DeePC_no_IVs}, as visualized by the dashed anti-diagonals.

\subsection{Noise mitigation using \acl{IVs}}
\noindent The \ac{CL-DeePC} formulation that was introduced in the previous subsection mitigates the closed-loop identification issue that arises from correlation between inputs and noise. However, sampling of noise in the past data matrix by the columns of $G$ in~\eqref{eq:CL_DeePC_no_IVs} may, similarly to \ac{DeePC}, still lead to predictions of unattainable of input-output trajectories~\citep{Markovsky2023}. This necessitates further noise mitigation strategies. The strategy that is presented here incorporates \ac{IVs} in a similar fashion as in~\citet{vanWingerden2022}. 

\subsubsection{Desirable properties of \ac{IVs}}\label{sec:IV_props}
\noindent As typically employed in estimation problems, \ac{IVs} are typically chosen to have two desirable properties that guarantee the consistency of the estimated quantity. Assuming sufficiently persistently exciting data such that $\Psi_{i,1,N}$ is full row rank, these conditions on the \acs{IV} matrix $\mathcal{Z}\in\mathbb{R}^{n_\mathrm{z}\times N}$ are~\citep{Verhaegen2007a}
\begin{align}
\text{rank}\Big(&\lim_{N\rightarrow\infty}\hat{\Sigma}_{\psi z}
\Big) = (p+1)r+pl, \label{eq:IV_preserve_info}\\
&\lim_{N\rightarrow\infty}\hat{\Sigma}_{ez}
\:=0,\label{eq:IV_noise_uncorr}
\end{align}
in which ${\hat{\Sigma}_{\psi z}=\Psi_{i,1,N}\mathcal{Z}^\top}$ and ${\hat{\Sigma}_{ez}=E_{i_p,1,N}\mathcal{Z}^\top}$ represent sample correlations that approach their respective true counterparts $\Sigma_{\psi z}$ and $\Sigma_{ez}$ in the limit $N\rightarrow\infty$. The former condition preserves the informativity of the employed data and the latter equation stipulates that the \acs{IV} is uncorrelated with noise.

\subsubsection{Assumptions}
\noindent Before moving on to the main results of this work, three assumptions are presented.
\setcounter{thm}{0}
\begin{assum}\label{assum:relative-rates}
    The past window length $p$ and the number of columns used to construct the block-Hankel data matrices $N$ go to infinity such that, with scalars $d,\alpha\in\mathbb{R}$,
    \begin{align}\label{eq:relative_rates}
        \begin{split}
            p &\geq -\frac{d\log N}{2\log|\rho|} \quad 1 < d < \infty,\\
            p&=o((\log N)^\alpha) \quad \alpha < \infty.
        \end{split}
    \end{align}
\end{assum}
For the system $\mathcal{S}$ from \secref{sec:sys_model} with $\rho<1$ this assumption is required for consistency of estimators that are based on closed-loop data because it ensures that errors due to mis-specification of the initial condition are $o_P(1/\sqrt{N})$ (and may therefore be neglected), and that sample correlations approach true correlations that are well-defined~\citep{Bauer2002,Chiuso2006}. This assumption will prove to be instrumental to demonstrate the closed-loop consistency of predictors\footnote{We shall refer to predictors as consistent if their bias (i.e. expected error) asymptotically goes to zero as $N\rightarrow\infty$.}.

\begin{assum}\label{assum:PE}
    The known past data matrix of the form $\Psi_{i,s,N}$, as defined by~\eqref{eq:Psi_def}, is full row rank.
\end{assum}
This assumption entails that used past input-output data is sufficiently persistently exciting, which is an important requirement for many data-driven applications~\citep{vanWaarde2023}.

\begin{assum}\label{assum:IV_def}
    The \acs{IV} matrix $\mathcal{Z}\in\mathbb{R}^{n_\mathrm{z}\times N}$ is chosen such that \eqref{eq:IV_preserve_info} and \eqref{eq:IV_noise_uncorr} hold.
\end{assum}
Under this assumption the chosen \acs{IV} matrix exhibits the desirable properties described in \secref{sec:IV_props}.

\subsubsection{Using \ac{IVs} and sequential step-ahead predictions}\label{sec:Theorem1}
\noindent This section presents results that pertain to the use of \ac{IVs} by \ac{CL-DeePC} to obtain a consistent predictor.
\setcounter{thm}{0}
\begin{lem}\label{lem:main_1}
    Consider input-output data generated (either in open or a well-posed closed-loop) by the minimal discrete \ac{LTI} system $\mathcal{S}$ from \secref{sec:sys_model}. By Assumption~\ref{assum:PE} the known data matrix $\Psi_{i,1,N}$ has full row rank. Furthermore, consider a partially known data matrix $\overline{\Psi}_{\hat{i},1,f}$ and output predictor $\widehat{Y}_{\hat{i}_p,1,f}^\mathrm{IV}$ as on the right hand side of Fig.~\ref{fig:CL-DeePC}, and an \acs{IV} matrix $\mathcal{Z}\in\mathbb{R}^{n_\mathrm{z}\times N}$ such that $\hat{\Sigma}_{\psi z}=\Psi_{i,1,N}\mathcal{Z}^\top$ is full row rank, 
    then $\exists G^\mathrm{IV}\in\mathbb{R}^{n_\mathrm{z}\times f}$ such that
    \begin{align}\label{eq:TheoremIV}
        \begin{bmatrix}
            \Psi_{i,1,N}\\Y_{i_p,1,N}
        \end{bmatrix}\mathcal{Z}^\top G^\mathrm{IV} =
        \begin{bmatrix}
            \overline{\Psi}_{\hat{i},1,f}\\\widehat{Y}_{\hat{i}_p,1,f}^\mathrm{IV}
        \end{bmatrix}.
    \end{align}
\end{lem}
\noindent\textit{Proof:} To start, note that any solution to the top row of \eqref{eq:TheoremIV} is consistent with the bottom row since $\widehat{Y}_{\hat{i}_p,1,f}^\mathrm{IV}$ is unknown a priori. In addition, since by assumption $\hat{\Sigma}_{\psi z}=\Psi_{i,1,N}\mathcal{Z}^\top$ is full row rank, based on the top row of \eqref{eq:TheoremIV}, solutions for $G^\mathrm{IV}$ are feasible and given by
\begin{align}\label{eq:G_sols}
    G^\mathrm{IV} = \hat{\Sigma}_{\psi z}^\dagger\overline{\Psi}_{\hat{i},1,f}+\hat{\Pi}_{\psi z}^\bot W,
\end{align}
where $\dagger$ indicates the Moore-Penrose pseudoinverse, ${\hat{\Pi}_{\psi z}^\bot=I_{n_\mathrm{z}}-\hat{\Sigma}_{\psi z}^\dagger\hat{\Sigma}_{\psi z}}$ is a matrix that forms an orthogonal projection onto the column space of $\hat{\Sigma}_{\psi z}$, and $W\in\mathbb{R}^{n_\mathrm{z}\times f}$ is a matrix of optimization variables. $\hfill\qed$

\setcounter{thm}{0}
\begin{rem}\label{rem:square_min_opt_var}
    Choosing $\mathcal{Z}$ such that $\hat{\Sigma}_{\psi z}=\Psi_{i,1,N}\mathcal{Z}^\top$ is square and invertible ensures that, with reference to \eqref{eq:G_sols}, $G^\mathrm{IV}$ is uniquely determined by $\overline{\Psi}_{\hat{i},1,f}$. Moreover, by \eqref{eq:TheoremIV}, $\widehat{Y}_{\hat{i}_p,1,f}^\mathrm{IV}$ is then also uniquely determined by $\overline{\Psi}_{\hat{i},1,f}$.
\end{rem}

Lemma~\ref{lem:main_1} effectively indicates that the system of equations provided by~\eqref{eq:TheoremIV} is consistent provided that $\hat{\Sigma}_{\psi z}$ is full row rank. The following theorem concerns the consistency of the resulting output predictor when using data that has been collected in closed-loop.
\setcounter{thm}{0}
\begin{thm}\label{theorem:main_result_IVs}
    Consider the minimal discrete \ac{LTI} system $\mathcal{S}$ from \secref{sec:sys_model} with $\rho<1$ to generate input-output data in a well-posed closed-loop configuration by means of a causal controller, known past data matrix $\Psi_{i,1,N}$ of full row rank by Assumption~\ref{assum:PE}, and partially-known data matrix $\overline{\Psi}_{\hat{i},1,f}$. Furthermore, let $p,N\rightarrow\infty$ as specified by Assumption~\ref{assum:relative-rates} and let ${\mathcal{Z}\in\mathbb{R}^{n_\mathrm{z}\times N}}$ satisfy Assumption~\ref{assum:IV_def}. Then the predictor $\widehat{Y}_{\hat{i}_p,1,f}^\mathrm{IV}$ specified by~\eqref{eq:TheoremIV} is consistent, i.e.:
    \begin{align}\label{eq:consistent}
    \lim_{p,N\rightarrow\infty}\mathbb{E}\left[\widehat{Y}_{\hat{i}_p,1,f}^\mathrm{IV}-Y_{\hat{i}_p,1,f}\right]=0.
    \end{align}
\end{thm}

\noindent \textit{Proof:} 
Starting with finite $p$ and $N$, by application of Lemma~\ref{lem:main_1}, \eqref{eq:TheoremIV} forms a consistent system of equations that specifies an output predictor $\widehat{Y}_{\hat{i}_p,1,f}^\mathrm{IV}$. Expanding the bottom row of~\eqref{eq:TheoremIV} by substitution of $Y_{i_p,1,f}$ as obtained from \eqref{eq:DataEq2} results in
\begin{align}\label{eq:Yf_hatIV_1}
\begin{split}\widehat{Y}_{\hat{i}_p,1,f}^\mathrm{IV} &= \widetilde{L}_1 \underbrace{\Psi_{i,1,N}\mathcal{Z}^\top G^\mathrm{IV}}_{=\overline{\Psi}_{\hat{i},1,f}} + \underbrace{E_{i_p,1,N}\mathcal{Z}^\top}_{=\hat{\Sigma}_{ez}} G^\mathrm{IV}\\ &\phantom{=}+ \widetilde{\Gamma}_1 \tilde{A}^p X_{i,1,N}\mathcal{Z}^\top G^\mathrm{IV},
\end{split}
\end{align}
wherein $\hat{\Sigma}_{ez}$ is recognized from its definition and $\overline{\Psi}_{\hat{i},1,f}$ is recognized from the top part of \eqref{eq:TheoremIV}. By comparison, from \eqref{eq:DataEq2}, the true output is
\begin{align}\label{eq:Yf_act}
    Y_{\hat{i}_p,1,f} &= \widetilde{L}_1 \Psi_{\hat{i},1,f} + E_{\hat{i}_p,1,f} + \widetilde{\Gamma}_1 \tilde{A}^p X_{\hat{i},1,f}.
\end{align}

Under Assumption~\ref{assum:relative-rates}, as $p,N\rightarrow\infty$ the sample correlations go to well-defined true correlations and the contribution of the initial condition may be neglected~\citep{Bauer2002,Chiuso2006}. In addition, by Assumption~\ref{assum:IV_def}, $\mathcal{Z}$ satisfies \eqref{eq:IV_noise_uncorr} such that it is uncorrelated with noise. Hence, computing the bias of the output predictor from \eqref{eq:Yf_hatIV_1} and \eqref{eq:Yf_act} in the aforementioned limits yields
\begin{align}\label{eq:asym_bias_IV}
    \lim_{p,N\rightarrow\infty}\!\!\mathbb{E}\left[\widehat{Y}_{\hat{i}_p,1,f}^\mathrm{IV}-Y_{\hat{i}_p,1,f}\right]\!=\!\!\lim_{p,N\rightarrow\infty}\!\!\widetilde{L}_1\mathbb{E}\left[\overline{\Psi}_{\hat{i},1,f}-\Psi_{\hat{i},1,f}\right]\!.
\end{align}
The remaining consistency proof for~\eqref{eq:consistent} is sequential. First, note the structure of $\overline{\Psi}_{\hat{i},1,f}$ and $\Psi_{\hat{i},1,f}$ as visualized by Fig~\ref{fig:CL-DeePC}. The leftmost columns of $\overline{\Psi}_{\hat{i},1,f}$ and $\Psi_{\hat{i},1,f}$ are equal because they contain no estimates. Therefore, the leftmost column of the predictor $\widehat{Y}_{\hat{i}_p,1,f}^\mathrm{IV}$ is consistent. Moreover, \eqref{eq:asym_bias_IV} shows that for subsequent columns in the output predictor the asymptotic bias is a linear combination of the asymptotic bias of preceding columns. Hence, since the leftmost column of the output predictor is consistent, so are the other columns of the output predictor, resulting in~\eqref{eq:consistent}. $\hfill\qed$

This theorem demonstrates that a suitable choice of $\mathcal{Z}$ facilitates consistent output predictions by means of~\eqref{eq:TheoremIV} when using noisy closed-loop data. The subsequent section presents such a suitable choice for $\mathcal{Z}$.

\subsubsection{Exemplary \acs{IV} that obtains a consistent predictor}\label{sec:example_IV}
\noindent The following lemma suggests a specific suitable \acs{IV} matrix for Theorem~\ref{theorem:main_result_IVs} that satisfies Assumption~\ref{assum:IV_def}, thus enabling consistent output predictions with~\eqref{eq:TheoremIV}.
\setcounter{thm}{1}
\begin{lem}\label{lem:example_IV}
For Theorem~\ref{theorem:main_result_IVs}, if the data-generating controller employs no direct feedthrough, a suitable choice of \acs{IV} matrix that satisfies Assumption~\ref{assum:IV_def} is $\mathcal{Z}=\Psi_{i,1,N}$.
\end{lem}
\textit{Proof:} This proof requires it to be shown that the choice ${\mathcal{Z}=\Psi_{i,1,N}}$ satisfies conditions \eqref{eq:IV_preserve_info} and \eqref{eq:IV_noise_uncorr} from Assumption~\ref{assum:IV_def} when the controller lacks direct feedthrough. Upon inspection, the rank condition specified by \eqref{eq:IV_preserve_info} is indeed satisfied by this choice because $\Psi_{i,1,N}$ is (by Assumption~\ref{assum:PE} in Theorem~\ref{theorem:main_result_IVs}) full row rank. With regards to \eqref{eq:IV_noise_uncorr}, substituting the choice of \acs{IV} and applying Assumption~\ref{assum:relative-rates} from Theorem~\ref{theorem:main_result_IVs} obtains
\begin{align}\label{eq:E_Phi_correlation}
    \!\!\!\!\begin{split}
        \lim_{p,N\rightarrow\infty}\!\!\hat{\Sigma}_{ez} &= \!\lim_{p,N\rightarrow\infty}\frac{1}{N}\;\sum\limits_{k=i_p}^{\mathclap{i_p+N-1}} e_k\!\begin{bmatrix}\datavec{u}{k-p,p+1}^\top & \datavec{y}{k-p,p}^\top \end{bmatrix}\\
        =\Sigma_{ez} &=\mathbb{E}\bigg[e_k\!\begin{bmatrix}\datavec{u}{k-p,p+1}^\top & \datavec{y}{k-p,p}^\top \end{bmatrix}\bigg]=0.
    \end{split}
\end{align}
The last equality in \eqref{eq:E_Phi_correlation} holds due to the (assumed) lack of direct feedthrough of the employed causal controller, due to which inputs are correlated with preceding noise (${\mathbb{E}[e_k u_j^\top]\neq0,\; \forall j>k}$), but inputs are uncorrelated with concurrent and subsequent noise (${\mathbb{E}[e_k u_j^\top]=0,\; \forall j\leq k}$). Moreover, the innovation noise is also uncorrelated with preceding outputs (${\mathbb{E}[e_k y_j^\top]=0,\; \forall j<k}$). $\hfill \qed$
\setcounter{thm}{1}
\begin{rem}
    For $G$ in \eqref{eq:CL_DeePC_no_IVs} consider the use of \ac{IVs} such that $G=\mathcal{Z}^\top G^\mathrm{IV}$ as in \eqref{eq:TheoremIV}. Based on~\eqref{eq:G_sols}, with full row rank $\Psi_{i,1,N}$, the choice $\mathcal{Z}=\Psi_{i,1,N}$ induces a minimum norm least squares solution for $G$.
\end{rem}

Lemma~\ref{lem:example_IV} suggests a choice of \acs{IV} matrix $\mathcal{Z}$ that ensures a consistent output predictor by Theorem~\ref{theorem:main_result_IVs}. With the sequential application of one-step-ahead predictors as per~\eqref{eq:TheoremIV}, the choice $\mathcal{Z}=\Psi_{i,1,N}$ is useful because it relies only on past input-output data, keeps the number of optimization variables in \eqref{eq:TheoremIV} to a minimum (see Remark~\ref{rem:square_min_opt_var}), and enforces causality of the predictor at finite~$N$. This last point will become clear in \secref{sec:Sequential}.

\subsection{Unified \ac{CL-DeePC} framework: incorporating \ac{IVs} and sequential multi-step-ahead predictors}
\noindent This section generalizes Theorem~\ref{theorem:main_result_IVs} to incorporate the sequential use of multi-step-ahead predictors. The corresponding generalization of \eqref{eq:TheoremIV} obtains the unified \ac{CL-DeePC} formulation
\begin{align}\label{eq:TheoremIV2}
    \begin{bmatrix}
        \Psi_{i,s,N}\\Y_{i_p,s,N}
    \end{bmatrix}\mathcal{Z}^\top G^\mathrm{IV} =
    \begin{bmatrix}
        \overline{\Psi}_{\hat{i},s,q}^\mathrm{m}\\\widehat{Y}_{\hat{i}_p,s,q}^\mathrm{IV,m}
    \end{bmatrix},
\end{align}
where $s$ is the multi-step-ahead predictor length, and $q$ is the number of sequential applications thereof, thus obtaining a total prediction length $f=sq$. The superscript $\mathrm{m}$ indicates that subsequent columns are shifted not by a single sample, as in Fig.~\ref{fig:CL-DeePC}, but by $s$ samples. Note that \ac{DeePC} with \ac{IVs} as in~\cite{vanWingerden2022} is recovered from \eqref{eq:TheoremIV2} by the single application ($q=1$) of a multi-step-ahead predictor of length $s=f$. For the unified formulation~\eqref{eq:TheoremIV2}, we present the following theorem.
\setcounter{thm}{1}
\begin{thm}\label{theorem:general_IV}
Consider the well-posed closed-loop system from Theorem~\ref{theorem:main_result_IVs} to generate input-output data, with $\Psi_{i,s,N}$ full row rank by Assumption~\ref{assum:PE} and partially-known data matrix $\overline{\Psi}_{i,s,q}^\mathrm{m}$. Furthermore, choose $\mathcal{Z}\in\mathbb{R}^{n_\mathrm{z}\times N}$ such that, with $p,N\rightarrow\infty$ as specified by Assumption~\ref{assum:relative-rates}, $\Psi_{i,s,N}\mathcal{Z}^\top$ is full row rank and $E_{i_p,s,N}\mathcal{Z}^\top\rightarrow0$. Then the predictor $\widehat{Y}_{\hat{i}_p,s,q}^\mathrm{IV,m}$ specified by \eqref{eq:TheoremIV2} is consistent, i.e.:
\begin{align*}
    \lim_{p,N\rightarrow\infty}\mathbb{E}\left[\widehat{Y}_{\hat{i}_p,s,q}^\mathrm{IV,m}-Y_{\hat{i}_p,s,q}\right]=0.
\end{align*}
\end{thm}
\textit{Proof:} 
Proof of this result follows the proof of Theorem~\ref{theorem:main_result_IVs}, only using a generic $s\in\mathbb{Z}_{>0}$ instead of $s=1$ and replacing the notation $f$ by $q$. $\hfill\qed$

Theorem~\ref{theorem:general_IV} presents the consistency of predictors obtained by the unified \ac{CL-DeePC} formulation~\eqref{eq:TheoremIV2}.

Suitable choices of the \ac{IV} matrix $\mathcal{Z}$ retain data informativity and are uncorrelated with noise as stipulated by Theorem~\ref{theorem:general_IV}. Note that with the choice $\mathcal{Z}=\Psi_{i,m,N}$, the condition $E_{i_p,s,N}\mathcal{Z}^\top\rightarrow0$ is violated for $m>1$, resulting in an inconsistent output predictor. Moreover, since $\Psi_{i,s,N}\mathcal{Z}^\top\in\mathbb{R}^{((p+s)r+pl)\times n_\mathrm{z}}$ must also be full row rank such that ${n_\mathrm{z}\geq(p+s)r+pl}$, this implies that for $s>1$ the \ac{IV} matrix $\mathcal{Z}$ must consist of more instruments than the past inputs and outputs contained by $\Psi_{i,1,N}$.

\setcounter{thm}{2}
\begin{rem}
Without \ac{IVs} (i.e. $\mathcal{Z}=I_N$), the use of a square and invertible data matrix ${\Psi_{i,s,N}\in\mathbb{R}^{((p+s)r+pl)\times N}}$ may seem appealing based on Remark~\ref{rem:square_min_opt_var}. However, the dimensions of a square data matrix $\Psi_{i,s,N}$ violate Assumption~\ref{assum:relative-rates}, which implies inconsistency of the predictor based on Theorem~\ref{theorem:general_IV}. This highlights the use of \ac{IVs} to obtain consistent output predictors.
\end{rem}

The special case with $q=f$ sequential applications of a one-step-ahead predictor ($s=1$) reduces~\eqref{eq:TheoremIV2} to~\eqref{eq:TheoremIV}, for which a computationally efficient implementation is presented in the subsequent section.
\section{Computationally efficient \acs{CL-DeePC}}\label{sec:Sequential}
\noindent This section presents an implementation of~\eqref{eq:TheoremIV} that reduces the number of optimization variables to improve the computational efficiency of \ac{CL-DeePC}, thus providing part of Contribution~4.

The use of such an efficient method can be understood by comparing the number of unknown, unequal optimization variables (in red) in Fig.~\ref{fig:CL-DeePC}, as summarized in Table~\ref{tab:unknowns} for the case with \ac{IVs}. Table~\ref{tab:unknowns} shows that for \ac{CL-DeePC} based on~\eqref{eq:TheoremIV} the influence of the number of unknowns in $G^\mathrm{IV}$ is quite large due to the relatively large minimum number of instruments $n_\mathrm{z}^\mathrm{min}$. To this end, this section will eliminate $G^\mathrm{IV}$ from the formulation provided by~\eqref{eq:TheoremIV}, considerably reducing the number of unknowns.

\begin{table}[t!]
    \centering
    \begin{tabular}{rrrcc}
        $G^\mathrm{IV}$ & inputs & outputs & total & $n_\mathrm{z}^\mathrm{min}$\\\hline
        $n_\mathrm{z}f$ & $fr$ & $fl$ & $f(r+l+n_\mathrm{z})$ & $p(r+l)+r$\\
    \end{tabular}
    \caption{Number of unequal unknowns in~\eqref{eq:TheoremIV} and their origin: $G^\mathrm{IV}$, inputs, and outputs. Full row rank of $\hat{\Sigma}_{\psi z}$ implies $n_\mathrm{z}\geq n_\mathrm{z}^\mathrm{min}$, which has a large influence on the total number of unknowns.}
    \label{tab:unknowns}
\end{table}

In line with Remark~\ref{rem:square_min_opt_var}, we choose $\mathcal{Z}\in\mathbb{R}^{n_\mathrm{z}\times N}$ with $n_\mathrm{z}=n_\mathrm{z}^\mathrm{min}$ from Table~\ref{tab:unknowns} such that $\hat{\Sigma}_{\psi z}$ is square and invertible, and consequently,  ${G^\mathrm{IV} = \hat{\Sigma}_{\psi z}\inv\overline{\Psi}_{\hat{i},1,f}}$ by~\eqref{eq:G_sols}. Substituting this result for $G^\mathrm{IV}$ in~\eqref{eq:TheoremIV} yields
\begin{align}\label{eq:Yfhat_1}
\widehat{Y}_{\hat{i},1,f}^\mathrm{IV}=\hat{\Sigma}_{yz}\hat{\Sigma}_{\psi z}\inv\overline{\Psi}_{\hat{i},1,f},
\end{align}
in which $\hat{\Sigma}_{yz}=Y_{i_p,1,f}\mathcal{Z}^\top$.

The sequential nature of \ac{CL-DeePC} can be exploited by the subsequent columns of $\overline{\Psi}_{\hat{i},1,f}$ in \eqref{eq:Yfhat_1}. For columns indexed by $k\in[\hat{i}_p,\hat{i}_p+f)$ this obtains
\begin{align}\label{eq:yhat_k}
    \hat{y}_{k} = \begin{bmatrix} \tilde{\beta}_1 & \cdots & \tilde{\beta}_{p+1} \end{bmatrix} \datavec{u}{k-p,p+1} + \begin{bmatrix} \tilde{\theta}_1 & \cdots & \tilde{\theta}_{p} \end{bmatrix} \datavec{\overline{y}}{k-p,p},
\end{align}
in which $\tilde{\beta}_j\in\mathbb{R}^{l\times r}$, and $\tilde{\theta}_j\in\mathbb{R}^{l\times l}$ are determined from $\hat{\Sigma}_{yz}\hat{\Sigma}_{\psi z}\inv$ in \eqref{eq:Yfhat_1}. Sequential application of the one-step-ahead predictor \eqref{eq:yhat_k} leads to
\begin{align}\label{eq:Sequential1}
    \datavec{\hat{y}}{\hat{i}_p,f} &=
    \begin{bmatrix}
        \widetilde{\mathcal{L}}^\mathrm{u}_f & \widetilde{\mathcal{G}}^\mathrm{u}_f 
    \end{bmatrix}    
    \begin{bmatrix}
        \datavec{u}{\hat{i},p}\\
        \datavec{u}{\hat{i}_p,f}
    \end{bmatrix}+
    \begin{bmatrix}
        \widetilde{\mathcal{L}}^\mathrm{y}_f & \widetilde{\mathcal{G}}^\mathrm{y}_f 
    \end{bmatrix}    
    \begin{bmatrix}
        \datavec{y}{\hat{i},p}\\
        \datavec{\hat{y}}{\hat{i}_p,f}
    \end{bmatrix},
\end{align}
in which
{\begingroup\allowdisplaybreaks
\begin{align*}
    \begin{bmatrix}
        \widetilde{\mathcal{L}}^\mathrm{u}_f & \widetilde{\mathcal{G}}^\mathrm{u}_f 
    \end{bmatrix}&= {\scriptsize
    \begin{bmatrix}
        \tilde{\beta}_1     & \cdots      & \tilde{\beta}_{p}   & \tilde{\beta}_{p+1} & 0           & 0           & \cdots      & 0          \\
        0           & \tilde{\beta}_1     & \cdots      & \tilde{\beta}_{p}   & \tilde{\beta}_{p+1} & 0           & \cdots      & 0          \\
        \vdots      & \ddots      & \ddots      &             & \ddots      & \ddots      & \ddots      & \vdots     \\
        0           & \cdots      & 0           & \tilde{\beta}_1     & \cdots      & \tilde{\beta}_{p}   & \tilde{\beta}_{p+1} & 0          \\
        0           & \cdots      & 0           & 0           & \tilde{\beta}_1     & \cdots      & \tilde{\beta}_{p}   & \tilde{\beta}_{p+1}\\
    \end{bmatrix}
    },\\
    \begin{bmatrix}
        \widetilde{\mathcal{L}}^\mathrm{y}_f & \widetilde{\mathcal{G}}^\mathrm{y}_f 
    \end{bmatrix}&= {\scriptsize
    \begin{bmatrix}
        \tilde{\theta}_1    & \cdots      & \tilde{\theta}_{p}  & 0            & 0            & 0           & \cdots       & 0          \\
        0           & \tilde{\theta}_1    & \cdots      & \tilde{\theta}_{p}   & 0            & 0           & \cdots       & 0          \\
        \vdots      & \ddots      & \ddots      &              & \ddots       & \ddots      & \ddots       & \vdots     \\
        0           & \cdots      & 0           & \tilde{\theta}_1     & \cdots       & \tilde{\theta}_{p}  & 0            & 0          \\
        0           & \cdots      & 0           & 0            & \tilde{\theta}_1     & \cdots      & \tilde{\theta}_{p}   & 0          \\
    \end{bmatrix}},
\end{align*} \endgroup}%
with ${\widetilde{\mathcal{L}}^\mathrm{u}_f\in\mathbb{R}^{fl\times pr}}$, ${\widetilde{\mathcal{G}}^\mathrm{u}_f\in\mathbb{R}^{fl\times fr}}$, ${\widetilde{\mathcal{L}}^\mathrm{y}_f\in\mathbb{R}^{fl\times pl}}$, and ${\widetilde{\mathcal{G}}^\mathrm{y}_f\in\mathbb{R}^{fl\times fl}}$. The subscript of these matrices indicates the number of block rows.

Notice that the predicted future outputs feature on both sides of \eqref{eq:Sequential1}. Solving for these outputs yields
\begin{align}\label{eq:Sequential2}
    \datavec{\hat{y}}{\hat{i}_p,f} &=
    \begin{bmatrix}
        \mathcal{L}^\mathrm{u}_f & \mathcal{L}^\mathrm{y}_f 
    \end{bmatrix}    
    \begin{bmatrix}
        \datavec{u}{\hat{i},p}\\
        \datavec{y}{\hat{i},p}
    \end{bmatrix}+
    \mathcal{G}^\mathrm{u}_f
    \datavec{u}{\hat{i}_p,f},
\end{align}%
in which $\mathcal{L}^\mathrm{u}_f$, $\mathcal{G}^\mathrm{u}_f$, and $\mathcal{L}^\mathrm{y}_f$ are uniquely defined by
\begin{align}\label{eq:Sequential3}
    \begin{bmatrix}
        \mathcal{L}^\mathrm{u}_f & \mathcal{G}^\mathrm{u}_f & \mathcal{L}^\mathrm{y}_f
    \end{bmatrix}=
    \left(I_{fl}-\widetilde{\mathcal{G}}^\mathrm{y}_f\right)\inv
    \begin{bmatrix}
        \widetilde{\mathcal{L}}^\mathrm{u}_f & \widetilde{\mathcal{G}}^\mathrm{u}_f & \widetilde{\mathcal{L}}^\mathrm{y}_f
    \end{bmatrix}.
\end{align}
The block-lower-triangular structure of $\widetilde{\mathcal{G}}^\mathrm{y}_f$ guarantees the invertibility of ${I_{fl}-\widetilde{\mathcal{G}}^\mathrm{y}_f}$ such that, \eqref{eq:Sequential3} can be solved directly to construct the predictor \eqref{eq:Sequential2}.

An efficient sequential procedure is also possible to solve \eqref{eq:Sequential3} that exploits the structure of ${I_{fl}-\widetilde{\mathcal{G}}^\mathrm{y}_f}$. For this sequential solution procedure, define the $f$ block-rows of $\big[\widetilde{\mathcal{L}}^\mathrm{u}_f \; \widetilde{\mathcal{G}}^\mathrm{u}_f \; \widetilde{\mathcal{L}}^\mathrm{y}_f\big]$ and $\big[\mathcal{L}^\mathrm{u}_f \; \mathcal{G}^\mathrm{u}_f \; \mathcal{L}^\mathrm{y}_f\big]$ by respectively $\tilde{\alpha}_j$, ${\alpha_j\in\mathbb{R}^{l\times p(r+l)+fr}}$, with $j$ here representing the index of the block row: $j=0,1,\dots,f-1$. It is then straightforward to show from \eqref{eq:Sequential3} that the formulation
\begin{align}\label{eq:Sequential4}
    \alpha_j=
    \left\{\begin{array}{ll}
    0          ,     & \text{if } j<0 \\
    \tilde{\alpha}_j,& \text{if } j=0 \\
    \tilde{\alpha}_j + \sum\limits_{m=1}^{p}\tilde{\theta}_m\alpha_{m-p+j-1}, & \text{if } j \geq 1
    \end{array}\right.
\end{align}
 allows efficient sequential construction of $\big[\mathcal{L}^\mathrm{u}_f \; \mathcal{G}^\mathrm{u}_f \; \mathcal{L}^\mathrm{y}_f\big]$ starting from $j=0$. 
 
 With reference to~\eqref{eq:Sequential1} we note that $\widetilde{\mathcal{G}}^\mathrm{u}_f$ is block-lower-triangular, just like ${I_{fl}-\widetilde{\mathcal{G}}^\mathrm{y}_f}$. Hence, by~\eqref{eq:Sequential3}, $\mathcal{G}^\mathrm{u}_f$ will also be block-lower-triangular. This block-lower-triangular structure enforces causality of the predictor~\eqref{eq:Sequential2}. This is unlike \ac{DeePC}, as shown by the fact that equivalent \ac{SPC} methods do not enforce causality of the predictor.

 In addition, from the subsequent section it will become clear that $\mathcal{G}^\mathrm{u}_f$ is not just block-lower-triangular, but also a block-Toeplitz matrix. This means that $\mathcal{G}^\mathrm{u}_f$ is fully parameterized by its leftmost block-column, allowing one to solve only for this part of $\mathcal{G}^\mathrm{u}_f$ in \eqref{eq:Sequential3} using \eqref{eq:Sequential4}. The complete matrix $\mathcal{G}^\mathrm{u}_f$ can then be constructed from its leftmost block-column thereafter.
\section{Equivalence to Closed-loop \acs{SPC}}\label{sec:equivalence2CLSPC}
\noindent This section demonstrates an equivalence between \ac{CL-DeePC} based on~\eqref{eq:TheoremIV} and the previous section and \ac{CL-SPC} as developed in~\cite{Dong2008}, thus providing the remainder of Contribution~4. The \ac{CL-SPC} algorithm is briefly explained first, based upon which the equivalence is demonstrated thereafter.

\subsection{Closed-loop \ac{SPC}}
\noindent To understand this equivalence, consider the data equations \eqref{eq:DataEq1} and \eqref{eq:DataEq2}. Just like \ac{CL-DeePC}, \ac{CL-SPC} uses $s=1$ to avoid closed-loop correlation between inputs and noise. \ac{CL-SPC} estimates the dynamic matrix $\widetilde{L}_1$ by least squares regression on past data:
\begin{align}\label{eq:CL-SPC-PredMarkov}
\hat{\widetilde{L}}_1 = \left[ \widehat{C\tKp{u}} \; \widehat{D} \; \widehat{C\tKp{y}} \right]=Y_{i_p,1,N}\mathcal{Z}^\top (\Psi_{i,1,N}\mathcal{Z}^\top)\inv,
\end{align}
in which typically, as in \secref{sec:example_IV}, $\mathcal{Z}=\Psi_{i,1,N}$. Assuming that $p$ is sufficiently large such that $\widetilde{A}^p=0$, estimates of the predictor Markov parameters contained in $\hat{\widetilde{L}}_1$ allow the construction of estimates of $\widetilde{\Gamma}_f\widetilde{\mathcal{K}}_p^\mathrm{u}$, $\widetilde{\Gamma}_f\widetilde{\mathcal{K}}_p^\mathrm{y}$ and $\widetilde{\mathcal{T}}_f^\mathrm{u}$, which make up $\widetilde{L}_f$, as well as $\tHf$. In line with~\eqref{eq:DataEq2} this allows the construction of a predictor as
\begin{align}\label{eq:CL-SPC-pred1}
	\begin{split}
	\datavec{\hat{y}}{\hat{i}_p,f}&= \begin{bmatrix}\widehat{\widetilde{\Gamma}_f\tKp{u}} & \widehat{\widetilde{\mathcal{T}}_f^\mathrm{u}} \end{bmatrix} 
	\begin{bmatrix}
		\datavec{u}{\hat{i},p}\\
		\datavec{u}{\hat{i}_p,f}
	\end{bmatrix}+\\
	&\phantom{=}\mkern8mu\begin{bmatrix}
		\widehat{\widetilde{\Gamma}_f\tKp{y}} & (I_{fl}-\widehat{\widetilde{\mathcal{H}}}_f) \end{bmatrix} 
	\begin{bmatrix}
		\datavec{y}{\hat{i},p}\\
		\datavec{\hat{y}}{\hat{i}_p,f}
	\end{bmatrix}.
	\end{split}
\end{align}
Note that the predicted output is contained on both sides of this equation. To solve \eqref{eq:CL-SPC-pred1} for the predicted outputs first make note of the fact that~\citep{Houtzager2012}
\begin{align}\label{eq:MatrixRelations}
	\begin{bmatrix}
		\Gamma_f\tKp{u} & \mathcal{T}_f^\mathrm{u} & \Gamma_f\tKp{y}
	\end{bmatrix}=
        \tHf\inv\begin{bmatrix}
		\widetilde{\Gamma}_f\tKp{u} & \widetilde{\mathcal{T}}_f^\mathrm{u} & \widetilde{\Gamma}_f\tKp{y}
	\end{bmatrix},
\end{align}
as is also visible from the combination of \eqref{eq:DataEq1} and \eqref{eq:DataEq2} for $s=f$. Using estimates in \eqref{eq:MatrixRelations} to solve \eqref{eq:CL-SPC-pred1} for the output predictions (which can be done efficiently in a sequential manner as with \ac{CL-DeePC}) yields
\begin{align}\label{eq:CL-SPC-pred2}
		\datavec{\hat{y}}{\hat{i}_p,f}= \begin{bmatrix}\widehat{\Gamma_f\tKp{u}} & \widehat{\Gamma_f\tKp{y}} \end{bmatrix} 
		\begin{bmatrix}
			\datavec{u}{\hat{i},p}\\
			\datavec{y}{\hat{i},p}
		\end{bmatrix}+
		\widehat{\mathcal{T}_f^\mathrm{u}} 
		\datavec{u}{\hat{i}_p,f},
\end{align}
which is in line with \eqref{eq:DataEq1} and can be used in a receding horizon optimal control framework.

\subsection{Equivalence between \ac{CL-DeePC} and \ac{CL-SPC}}
\noindent By comparing \eqref{eq:Sequential1} to \eqref{eq:CL-SPC-pred1} for $f=1$ it can be seen that the building blocks $\tilde{\beta}_j$ and $\tilde{\theta}_j$ in \ac{CL-DeePC} are equal to the estimated predictor Markov parameters from \eqref{eq:CL-SPC-PredMarkov} that are obtained using \ac{CL-SPC}:
\begin{subequations}\label{eq:EquivMarkov}
	\begin{align}
		\tilde{\beta}_j &=
            \left\{%
            \begin{array}{ll}
			\widehat{C\tilde{A}^{p-j}\tilde{B}} &\qquad 1 \leq j \leq p,\\
			\widehat{D} &\qquad j=p+1,
		\end{array}
            \right.\label{eq:EquivMarkovInputs}\\
	\tilde{\theta}_j &= \widehat{C\tilde{A}^{p-j}K} \mkern24mu \qquad 1 \leq j \leq p
        \label{eq:EquivMarkovOutputs}.
	\end{align}
\end{subequations}
As a result there is an equivalence between the constructed matrices in \ac{CL-SPC} and the efficient sequential \ac{CL-DeePC} method that is succinctly described by
\begin{align}\label{eq:EquivPredictorMatrices}
	\mkern-3mu\begin{bmatrix}
		\widetilde{\mathcal{L}}^\mathrm{u}_f & \widetilde{\mathcal{G}}^\mathrm{u}_f & \widetilde{\mathcal{L}}^\mathrm{y}_f & \left(I_{fl}-\widetilde{\mathcal{G}}^\mathrm{y}_f\right)
	\end{bmatrix}\mkern-3mu=\mkern-3mu\begin{bmatrix}
		\widehat{\widetilde{\Gamma}_f\tKp{u}} & \widehat{\widetilde{\mathcal{T}}_f^\mathrm{u}} & \widehat{\widetilde{\Gamma}_f\tKp{y}} & \widehat{\widetilde{\mathcal{H}}}_f
	\end{bmatrix}\mkern-3mu.
\end{align}
It follows from \eqref{eq:Sequential3}, \eqref{eq:MatrixRelations}, and \eqref{eq:EquivPredictorMatrices} that, likewise,
\begin{align}\label{eq:EquivInnovationMatrices}
	\begin{bmatrix}
		\mathcal{L}^\mathrm{u}_f & \mathcal{G}^\mathrm{u}_f & \mathcal{L}^\mathrm{y}_f
	\end{bmatrix}=\begin{bmatrix}
		\widehat{\Gamma_f\tKp{u}} & \widehat{\mathcal{T}_f^\mathrm{u}} & \widehat{\Gamma_f\tKp{y}}
	\end{bmatrix}.
\end{align}
Moreover,~\eqref{eq:EquivInnovationMatrices} demonstrates that the output predictors~\eqref{eq:Sequential2} and~\eqref{eq:CL-SPC-pred2} of respectively \ac{CL-DeePC} and \ac{CL-SPC} are equivalent. By comparison of the sequential technique to solve \eqref{eq:Sequential3} using \eqref{eq:Sequential4} to the sequential algorithm of~\cite{Dong2008} it is furthermore possible to see that the two sequential algorithms are indeed equivalent\footnote{Note that direct-feedthrough is not considered in~\cite{Dong2008}, but that an expansion of $B=\tilde{B}+KD$ would yield an algorithm consistent with \ac{CL-DeePC} and~\eqref{eq:MatrixRelations}.}.

\section{Results}\label{sec:results}
\noindent This section presents simulation results of \ac{DeePC} and \ac{CL-DeePC} that facilitate the comparison of their performance, thereby providing Contribution~5. We will first describe the simulation setup before presenting a reference tracking example. A subsequent demonstration of closed-loop correlation between inputs and noise is followed by a consistency analysis. Lastly, a parametric sensitivity analysis is performed.
\subsection{Simulation setup}\label{sec:simulation_setup}
\noindent This section presents the simulation setup. The simulated discrete plant is a marginally stable fifth-order 
model of the form \eqref{eqn:SS_innovation} that represents two circular plates that are spun by a motor with non-rigid shafts with \citep{Favoreel1999b}
\begin{align}\label{eq:SysFavoreel}
\scriptsize
\begin{array}{lll}
    A = \begin{bmatrix}
        4.40 & 1 & 0 & 0 & 0\\
       -8.09 & 0 & 1 & 0 & 0\\
        7.83 & 0 & 0 & 1 & 0\\
       -4.00 & 0 & 0 & 0 & 1\\
        0.86 & 0 & 0 & 0 & 0
    \end{bmatrix}\!,&
    B = \begin{bmatrix}
        0.00098\\
        0.01299\\
        0.01859\\
        0.0033\\
       -0.00002
    \end{bmatrix}\!,&C = \begin{bmatrix}
        1 \\ 0 \\ 0\\ 0\\ 0
    \end{bmatrix}^\top\mkern-7mu,\\
    K = \begin{bmatrix}
        2.3 & -6.64 & 7.515 & -4.0146 & 0.86336
    \end{bmatrix}^\top\mkern-7mu,\span
    &D=0.
\end{array}
\end{align}

Three different controllers are implemented. One controller uses \ac{DeePC} with \ac{IVs} as proposed in \cite{vanWingerden2022}\footnote{The implementation relies on an equivalent, but computationally more efficient \ac{SPC} formulation.}. Another uses \ac{CL-DeePC} with \ac{IVs}, implemented as described in \secref{sec:Sequential} to reduce computation time. As a benchmark, an oracle model predictive controller based on \eqref{eq:SS_iter} with perfect knowledge of the system matrices $\Gamma_f$, $\mathcal{T}_f^\mathrm{u}$ and state $x_{\hat{i}_p}$ is used.

All of the simulations are carried out in MATLAB\footnote{The employed code and resulting data can respectively be found via doi.org/10.5281/zenodo.10573259 and doi.org/10.5281/zenodo.10573874.}, using CasADi \citep{Andersson2019} to formulate a quadratic program that is solved with IPOPT \citep{Wachter2006}. Unless indicated otherwise $p=f=20$, the employed cost function weights as in \eqref{eq:cost_fun} are $Q=100$, $R=0$, and $R^\Delta=10$, and employed constraints are $|u_k-u_{k-1}|\leq3.75$, $|u_k|\leq15$, $|y_k|<1000$. To prevent infeasibility these constraints are relaxed using quadratically penalized slack variables (with weighting $\lambda=10^{15}$) and the problem is solved again if no solution is found. This appears to be necessary with \ac{DeePC}, particularly for $\bar{N}<200$.

All simulations use a square wave reference signal that varies between 0 and 100 with a period of 200 time steps. Simulations start by data collection in open-loop with which to initialize the data-driven controllers, which are then active together with the oracle for 1800 time steps. Unless otherwise stated, the input variance for the open-loop trajectories is $\Sigma(u_k)=1$ and the innovation noise has a variance of $\Sigma(e_k)=1$ throughout.
\subsection{Example of a reference tracking case}
\begin{figure}[b!]
\begin{center}
\includegraphics[width=\columnwidth]{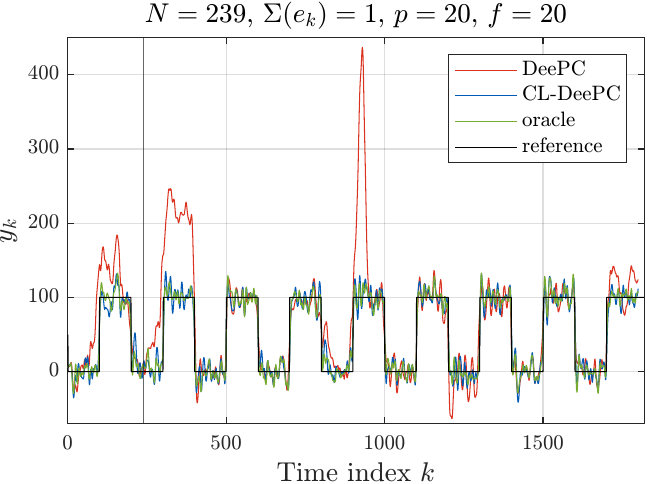}
\caption{Reference tracking by adaptive \ac{DeePC} and \ac{CL-DeePC} using \ac{IVs}. After the vertical line at $\bar{N}$ all used data originates from operation in closed-loop. \ac{CL-DeePC} displays better reference tracking performance than \ac{DeePC}, in part due to a closed-loop identification problem.}
\label{fig:CL_Problem_Solution}
\end{center}
\end{figure}

\noindent To compare the performance of \ac{DeePC} and \ac{CL-DeePC} in an adaptive setting this section presents a reference tracking example by Fig.~\ref{fig:CL_Problem_Solution}. \ac{CL-DeePC} clearly manages considerably better reference tracking performance w.r.t. \ac{DeePC}, the former performing comparably to the oracle. At first, all of the data that the data-driven controllers rely on derives from open-loop operation. Due to the adaptive nature of the controller implementation, after $\bar{N}$ time steps, all of the employed data derives from closed-loop operation. The reference tracking ability of \ac{DeePC} appears to decrease as the amount of employed closed-loop data increases. Thereafter, a cyclical behaviour may be observed for \ac{DeePC}: large reference tracking errors that result from, e.g., the closed-loop identification issue momentarily increase the signal-to-noise ratio and thereby temporarily improve the obtained reference tracking performance again.

\subsubsection{Correlation between inputs and noise}
\begin{figure}[b!]
\begin{center}
\includegraphics[width=\columnwidth]{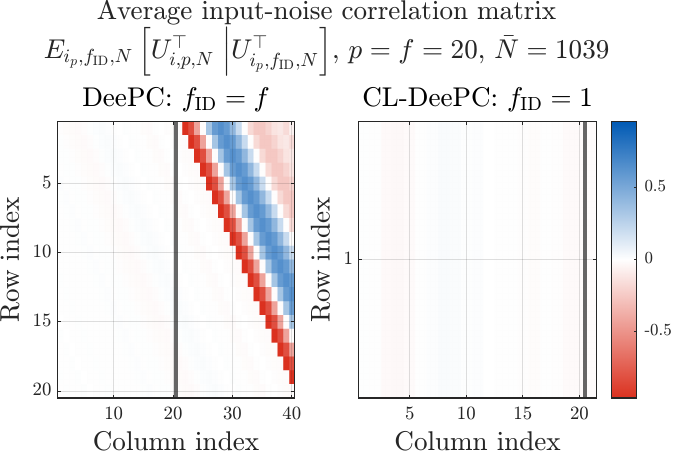}
\caption{Noise-input correlation matrix for \ac{DeePC} and \ac{CL-DeePC} averaged over the closed-loop data from 120 noise realizations. In contrast to \ac{DeePC}, \ac{CL-DeePC} makes use of input data that is uncorrelated with preceding noise.}  
\label{fig:EfUpf_correlation}                                 
\end{center}                                 
\end{figure}

\noindent This section demonstrates the existence of correlation between inputs and preceding noise during closed-loop operation in an adaptive setting by means of simulations. Following the preceding consistency proof in \secref{sec:Theorem1}, the correlation matrix of interest is given by $E_{i_p,f_\mathrm{ID},N}\begin{bmatrix}U_{i,p,N}^\top & U_{i_p,f_\mathrm{ID},N}^\top\end{bmatrix}$, with $f_\mathrm{ID}=f$ for \ac{DeePC}, and $f_\mathrm{ID}=1$ for \ac{CL-DeePC}. Fig.~\ref{fig:EfUpf_correlation} shows the average of this correlation matrix based on the closed-loop data of 120 different noise realizations. Based on the figure, between \ac{DeePC} and \ac{CL-DeePC}, only \ac{DeePC} experiences the correlation that it induces between inputs and noise. Note that the stochastic variability of subsequent control policies, which arises from the adaptive implementation in the presence of noise, is insufficient to mitigate the input-noise correlation experienced by \ac{DeePC}.

\subsubsection{Consistency analysis}
\begin{figure}[b!]
\begin{center}
\includegraphics[width=\columnwidth]{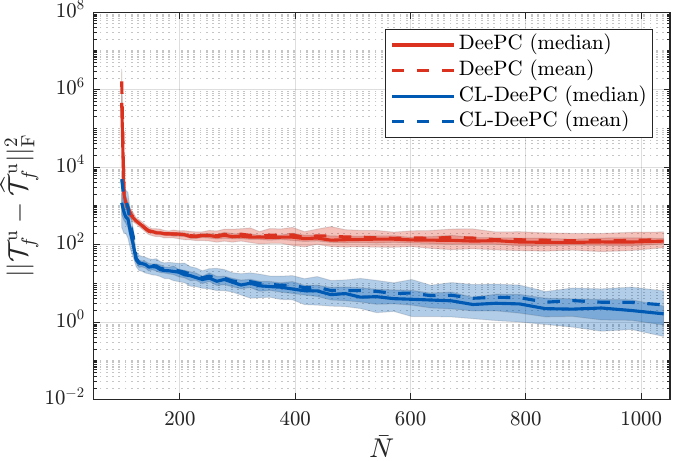}
\caption{Bias of the implicitly estimated matrix $\mathcal{T}_f^\mathrm{u}$ based exclusively on adaptive closed-loop operation. Shaded regions indicate the 10\textsuperscript{th}, 30\textsuperscript{th}, 70\textsuperscript{th} and 90\textsuperscript{th} percentiles of 120 simulations with different noise realizations.}
\label{fig:Tuf_consistency}                                 
\end{center}                                 
\end{figure}

\noindent This section demonstrates the consistency (or lack thereof) of the estimators employed by \ac{DeePC} and \ac{CL-DeePC} as a result of experienced closed-loop input-noise correlation. Along the lines of the consistency analysis in \secref{sec:Theorem1} and~\cite{Dinkla2023} it is to be expected that the implicit matrix estimate of $\mathcal{T}_{f_\mathrm{ID}}^\mathrm{u}$ is inconsistent for \ac{DeePC} ($f_\mathrm{ID}=f$) and consistent for \ac{CL-DeePC} ($f_\mathrm{ID}=1$). For a fair comparison, the error in $\widehat{\mathcal{T}}_f^\mathrm{u}$ is shown for both controllers by Fig.~\ref{fig:Tuf_consistency}. \secref{sec:Sequential} is followed to construct this estimate for \ac{CL-DeePC}.

As expected, as the number of employed past data points $\bar{N}$ increases (together with the number of columns $N=\bar{N}-p-f_\mathrm{ID}+1$), the bias of the \ac{CL-DeePC} estimate keeps decreasing, which indicates that the employed estimate is indeed consistent. In contrast, for \ac{DeePC} the bias does not keep decreasing noticeably beyond around $\bar{N}=400$, indicating that the implicitly employed estimate $\widehat{\mathcal{T}}_f^\mathrm{u}$ is inconsistent.
\subsection{Parametric sensitivity analysis}\label{sec:sensitivity_study}
\noindent This section performs a parametric sensitivity study on the reference tracking performance of \ac{DeePC}, \ac{CL-DeePC}, and the oracle. Investigated parameters are the number of past data points $\bar{N}$, the innovation noise variance $\Sigma(e_k)$, and the window lengths $p=f$. As a measure of the reference tracking performance a scaled root mean square of the reference tracking error is used: $J_\mathrm{rms}=\sqrt{\big(\sum_k r_k^2\big)\inv\sum_k(y_k-r_k)^2}$. To reflect purely the effect on performance during adaptive closed-loop operation, the first $\bar{N}$ control actions are excluded. Note that it is possible for \ac{DeePC} and \ac{CL-DeePC} to outperform the oracle because the oracle does also not account for noise.

\subsubsection{Number of past data samples: $\bar{N}$}
\begin{figure}[t!]
\begin{center}
\includegraphics[width=\columnwidth]{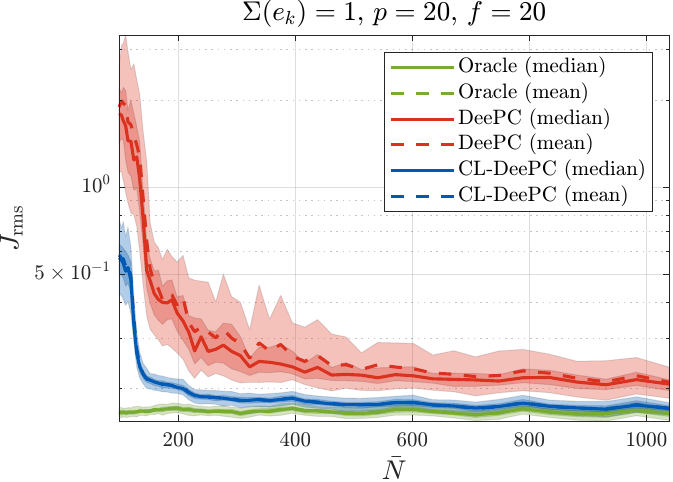}
\caption{Effect of $\bar{N}$ on reference tracking performance. Shaded regions indicate the 10\textsuperscript{th}, 30\textsuperscript{th}, 70\textsuperscript{th} and 90\textsuperscript{th} percentiles of 120 simulations.\\}  
\label{fig:varying_Nbar}                                 
\end{center}                                 
\end{figure}

\noindent The effect of a varying number of past data samples $\bar{N}$ is shown in Fig.~\ref{fig:varying_Nbar}. Slight fluctuations of the displayed oracle performance are an artifact that is attributable to the exclusion of the first $\bar{N}$ control actions in the calculation of $J_\mathrm{rms}$ to exclude control actions that are based on open-loop data. Using \ac{CL-DeePC} the reference tracking performance approaches the oracle's performance at a relatively low $\bar{N}$. For \ac{DeePC} the picture is different, with the median reference tracking performance remaining $26\%$ higher than its oracle counterpart at $\bar{N}=1039$. By comparison this value is only $4\%$ for \ac{CL-DeePC}.

Fig.~\ref{fig:varying_Nbar} demonstrates that compared to \ac{DeePC}, in general \ac{CL-DeePC} is able to use fewer past data samples to obtain as good or better performance, hence refered to as better sample efficiency. Compared to \ac{DeePC}, the better sample efficiency of \ac{CL-DeePC} is attributable to the use of a shorter future prediction window length for identification ($f_\mathrm{ID}=1$ as opposed to $f_\mathrm{ID}=f$). This leaves more columns $N=\bar{N}-p-f_\mathrm{ID}+1$ to approximate the relevant correlation matrices that are used implicitly by both \ac{DeePC} and \ac{CL-DeePC}. In addition, the discussed closed-loop identification issue entails that even if $\bar{N}$, and therefore $N$, is large such that these correlation matrices are approximated well, \ac{CL-DeePC} outperforms \ac{DeePC}.

\subsubsection{Noise level: $\Sigma(e_k)$}
\begin{figure}[t!]
\begin{center}
\includegraphics[width=\columnwidth]{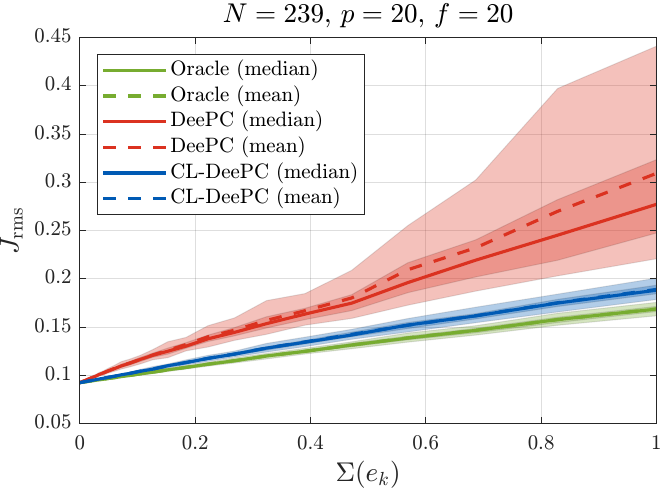}
\caption{Effect of $\Sigma(e_k)$ on reference tracking performance. Shaded regions indicate the 10\textsuperscript{th}, 30\textsuperscript{th}, 70\textsuperscript{th} and 90\textsuperscript{th} percentiles over 120 simulations.}  
\label{fig:varying_Re}                                 
\end{center}                                 
\end{figure}

\noindent The effect of the noise level, as quantified by a varying innovation noise variance $\Sigma(e_k)$, is shown in Fig.~\ref{fig:varying_Re}. In the absence of noise, \ac{DeePC} and \ac{MPC} are equivalent~\citep{Coulson2019}. In the noiseless case, the closed-loop identification issue does not arise so the performance of the three algorithms is identical. As the noise level increases, correlation between inputs and preceding noise increases because $|e_k|$ is typically larger, and more control effort is needed to perform noise rejection. Consequently, the closed-loop identification issue becomes more troublesome for \ac{DeePC} at higher noise levels. Note that the performance of all methods decreases with an increasing noise level because all methods lack the ability to immediately compensate noise disturbances. Using the slopes of the median performance as a measure for the susceptibility to noise-induced performance deterioration, \ac{CL-DeePC} is 48\% less susceptible to noise-induced performance deterioration compared to \ac{DeePC}.

\subsubsection{Data window lengths: $p=f$}
\begin{figure}[t!]
\begin{center}
\includegraphics[width=\columnwidth]{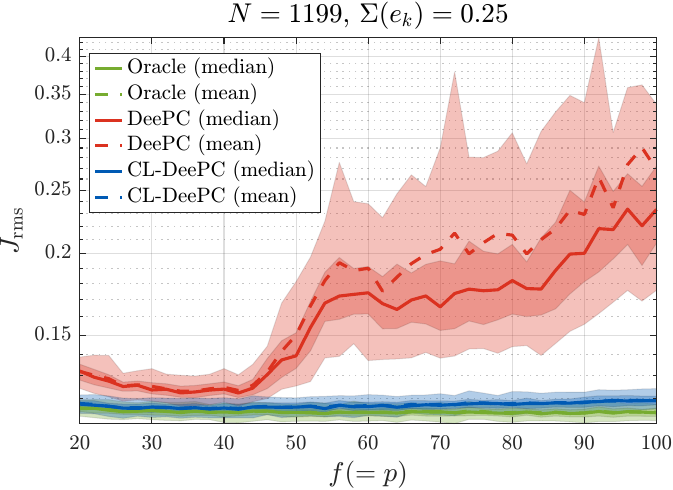}
\caption{Effect of $f=p$ on reference tracking performance. Shaded regions indicate the 10\textsuperscript{th}, 30\textsuperscript{th}, 70\textsuperscript{th} and 90\textsuperscript{th} percentiles of 120 simulations.}  
\label{fig:varying_pf}                                 
\end{center}                                 
\end{figure}
\noindent This section investigates the effect of the window lengths with $p=f$. The choice $p=f$ is made for the sake of simplicity is further motivated by its frequent use in the closely related field of subspace identification~\citep{vanderVeen2013}. We discern two competing mechanisms to explain the results shown in Fig.~\ref{fig:varying_pf} for \ac{CL-DeePC} and an additional detrimental effect on performance for \ac{DeePC}. Increasing $p$ is beneficial in terms of reducing the bias of the predictor. However, as with increasing $f_\mathrm{ID}$, increasing $p$ entails implicitly estimating more parameters of the predictor, thereby increasing its variance. This is more pronounced for \ac{DeePC} since $f_\mathrm{ID}=f$ when compared to \ac{CL-DeePC} for which $f_\mathrm{ID}=1$. The shorter future window length used for identification $f_\mathrm{ID}$ is also beneficial for \ac{CL-DeePC} because a higher degree of collinearity may be expected for larger $f_\mathrm{ID}$, potentially leading to ill-conditioning of the identification task~\citep{Chiuso2004}. Together with the aforementioned closed-loop identification issue these effects induce up to a 49\% lower reference tracking cost at $p=f=100$ for \ac{CL-DeePC} compared to \ac{DeePC}. Moreover, note that the performance of \ac{DeePC} deteriorates rapidly with increasing $p=f$, whereas the impact on the performance of \ac{CL-DeePC} is far less pronounced.

\section{Conclusion}\label{sec:conclusion}
\noindent This article establishes \ac{CL-DeePC} to address a closed-loop identification issue that is inherent to adaptive \ac{DeePC} implementations. By incorporating \ac{IVs} to mitigate noise it is shown that \ac{CL-DeePC} employs a consistent output predictor when relying on data obtained in closed-loop. Moreover, to limit the number of optimization variables, an efficient sequential procedure is established to construct the output predictor in practice. This sequential procedure reveals the equivalence of the developed \ac{CL-DeePC} method and \ac{CL-SPC}.

Simulations confirm that \ac{CL-DeePC} relies on data that does not exhibit correlation between inputs and preceding noise, facilitating the use of a consistent output predictor. Furthermore simulations demonstrate superior reference tracking performance of \ac{CL-DeePC} compared to \ac{DeePC}. A sensitivity analysis illustrates that \ac{CL-DeePC} is more sample efficient than \ac{DeePC}, less sensitive to high noise levels, and performs better for a wide range of window lengths $p=f$. Where this work has considered a single specific choice of \acs{IV} ($\mathcal{Z}=\Psi_{i,1,N}$), future work may consider different \ac{IVs} in \ac{CL-DeePC} to further reduce the effects of noise. Moreover, where simulations have employed the computationally efficient implementation of~\eqref{eq:TheoremIV}, future work may further explore the merits of different settings (not $s=1$ and $q=f$) of the unified \ac{CL-DeePC} approach provided by~\eqref{eq:TheoremIV2}.

\bibliographystyle{elsarticle-harv}
\bibliography{autosam}           

\appendix
\end{document}